\documentclass[aoas,nameyear,seceqn,dvips]{arximspdf}
\usepackage{graphicx}

\doi{10.1214/10-AOAS434}
\volume{5}
\issue{2A}
\pubyear{2011}
\firstpage{605}
\lastpage{627}

\begin{document}
\begin{frontmatter}

\title{Analysis of rolling group therapy data using conditionally autoregressive priors}
\runtitle{Car priors for rolling group therapy data}
\vspace*{-3pt}
\begin{aug}
\author[a]{\fnms{Susan M.} \snm{Paddock}\corref{}\thanksref{t1,t2}\ead[label=e1]{paddock@rand.org}},
\author[a]{\fnms{Sarah B.} \snm{Hunter}\thanksref{t1,t3}\ead[label=e2]{shunter@rand.org}},
\author[a]{\fnms{Katherine~E.}~\snm{Watkins}\thanksref{t1,t3}\ead[label=e3]{kwatkins@rand.org}}
\and
\author[b]{\fnms{Daniel F.} \snm{McCaffrey}\thanksref{t1}\ead[label=e4]{danielm@rand.org}
\ead[label=u1,url]{http://www.rand.org}}
\thankstext{t1}{Supported in part by NIH/NIAAA Grant R01AA019663.}
\thankstext{t2}{Supported in part by NIH/NIDDK Grant R01DK061662.}
\thankstext{t3}{Supported in part by NIH/NIAAA Grant R01AA014699.}
\runauthor{Paddock, Hunter, Watkins and McCaffrey}

\affiliation{RAND Corporation}

\address[a]{S. A. Paddock\\
S. B. Hunter\\
K. E. Watkins\\
RAND Corporation\\
1776 Main Street\\
Santa Monica, California 90401\\USA\\
\printead{e1}\\
\phantom{E-mail:\ }\printead*{e2}\\
\phantom{E-mail:\ }\printead*{e3}}

\address[b]{D. F. McCaffrey\\
RAND Corporation\\
4570 Fifth Avenue, Suite 600\\
Pittsburgh, Pennsylvania 15213\\USA\\
\printead{e4}}
\end{aug}

\received{\smonth{7} \syear{2010}}

\begin{abstract}
Group therapy is a central treatment modality for behavioral
health disorders such as alcohol and other drug use (AOD) and
depression. Group therapy is often delivered under a rolling
(or open) admissions policy, where new clients are continuously
enrolled into a~group as space permits. Rolling admissions
policies result in a complex correlation structure among client
outcomes. Despite the ubiquity of rolling admissions in
practice, little guidance on the analysis of such data is
available. We discuss the limitations of previously proposed
approaches in the context of a study that delivered group
cognitive behavioral therapy for depression to clients in
residential substance abuse treatment. We improve upon previous
rolling group analytic approaches by fully modeling the
interrelatedness of client depressive symptom scores using a
hierarchical Bayesian model that assumes a conditionally
autoregressive prior for session-level random effects. We
demonstrate improved performance using our method for
estimating the variance of model parameters and the enhanced
ability to learn about the complex correlation structure among
participants in rolling therapy groups. Our approach broadly
applies to any group therapy setting where groups have changing
client composition. It will lead to more efficient analyses of
client-level data and improve the group therapy research
community's ability to understand how the dynamics of rolling
groups lead to client outcomes.\vspace*{-3pt}
\end{abstract}


\end{frontmatter}

\section{Introduction}

\subsection{Group therapy}

Group therapy is a central treatment modality for behavioral
health disorders, such as alcohol and other drug use (AOD)
disorders [\citet{kadd2001}; \citet{kad2004}; \citet{mon2002}; \citet{crit999}; \citet{well1994}]
and depression [\citet{tho1983}; \citet{neiberm1989}; \citet{robi1990}; \citet{brig1999}].
Most community-based behavioral health treatment is provided in
groups. Its therapeutic strengths include group members having
opportunities to develop and practice new social skills and
behaviors, receiving feedback from other group members,
learning from the shared experiences of other group members,
and providing a recovery network to clients
[\citet{neiberm1989}; \citet{satt1994}]. Group therapy is
especially relevant for treatment providers in light of rising
costs, as group therapy has clear economic advantages over
individual therapy [\citet{jac1989}; \citet{brig1999}; \citet{mon2002}]. These advantages are evident in our
motivating study, Building Recovery by Improving Goals, Habits
and Thoughts (BRIGHT). In BRIGHT, AOD treatment clients with
depressive symptoms received group cognitive behavioral therapy
(CBT) that was delivered by AOD treatment counselors, rather
than psychotherapists. The analytic question we consider in
this paper is whether the depressive symptoms of AOD clients
decreased while attending group CBT over an 8-week period.

\begin{figure*}[b]

\includegraphics{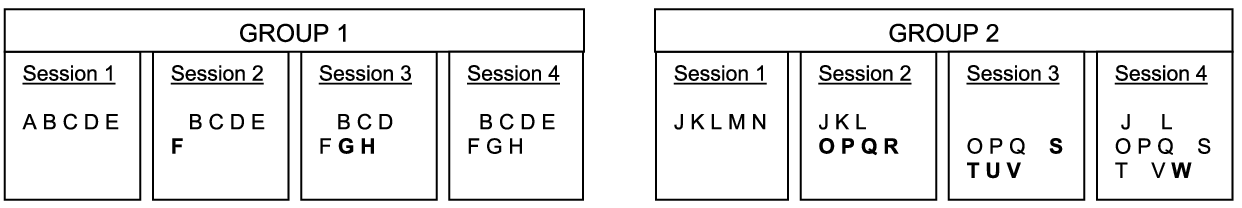}

\caption{Illustration of client flow under rolling admissions.
Each group contains different sets of clients (represented by letters).
Bold letters represent new members entering each group.}
\label{fig1a}
\end{figure*}

\subsection{Group therapy with rolling admissions}

A feature of the group CBT offered in BRIGHT that is shared by
many therapy groups offered in other AOD treatment settings is that
clients enter group under a rolling (or open) admissions
policy. This means that clients are continuously enrolled in
group as space permits. Figure~\ref{fig1a} illustrates a
hypothetical rolling admission policy for two therapy groups
(labeled ``GROUP 1'' and ``GROUP 2''), each of which has four
sessions.  It is important to note the difference between {\it
groups} and \textit{sessions}. The two groups in Figure~\ref{fig1a}
each have four sessions. The sessions in Group 1 are
independent of sessions in Group 2 because there is no overlap
in the sets of clients who attend the sessions in Group 1
versus Group 2---for example, in Figure~\ref{fig1a}, clients A--H
attend Group 1 and clients J--W attend Group 2. In contrast, the
sessions within each group are not independent since clients
may attend multiple sessions within each group. For example,
clients B, C, D attend all four sessions offered in Group~1.
The ``rolling'' admissions policy is depicted in
Figure~\ref{fig1a}, with new members indicated by boldface font
in the first session they attend. While new members join the
group, other members leave (e.g., client A left Group 1 after
session~1 and was replaced by client~F in session 2). This
structure induces a complex pattern of interrelatedness among
clients. Even though clients A and~F in Group 1 never attend
the same session, their data could be correlated. One
example of this would be if client~A were a disruptive client
[\citet{csat2005}] who adversely influenced the overall group
dynamic in session~1 of Group 1 in such a way that clients B
through~E would bring that negativity with them to session 2,
thus affecting client~F's group experience.

\begin{figure*}

\includegraphics{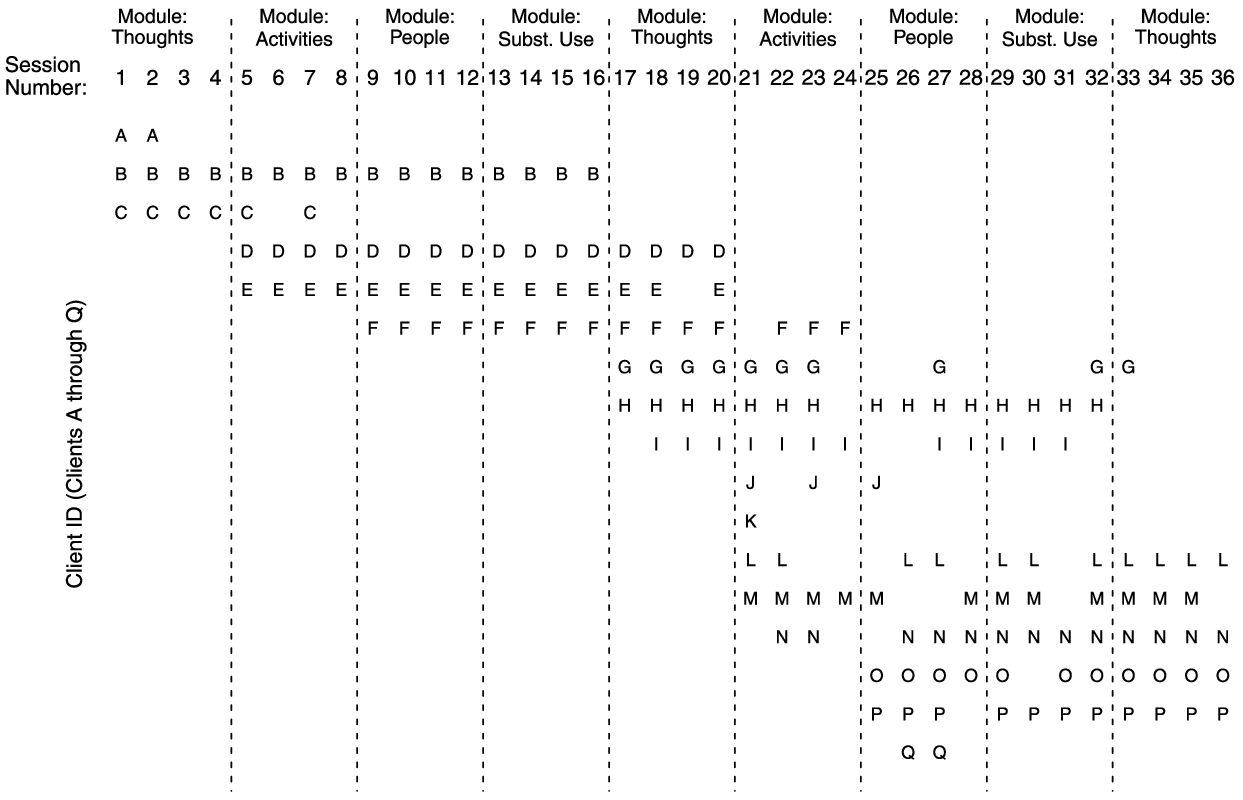}%
\vspace*{-3pt}
\caption{Attendance by clients A--Q for sessions 1--36 in one rolling therapy group in the BRIGHT study.
Study modules are separated by dashed vertical lines.}
\label{fig1b}
\vspace*{-3pt}
\end{figure*}

Figure~\ref{fig1b} illustrates the actual client attendance
patterns for $17$ clients who participated in one of the four
rolling cognitive behavioral therapy (CBT) groups in the BRIGHT
study. This group included $36$ sessions. Each client (labeled
A through Q) was expected to complete $16$ sessions, with $4$
sessions coming from each of four modules (or session themes)---\textit{Thoughts}, \textit{Activities}, \textit{People},
or \textit{Substance Abuse}---as noted at the top of Figure~\ref{fig1b}. Modules were
offered on a rotating basis, and clients were allowed to enter
the group at the start of a new module (e.g., at session 5, 9,
13, 17, 21, 25, or 33). This illustrates the flexibility of
treatment delivery afforded by rolling groups in BRIGHT---for example, a client only needed
to wait for a new module, rather
than an entirely new but closed 16-session group, to start CBT.

Thus, not only are client outcomes correlated \textit{within}
therapy sessions, but they are also correlated \textit{across}
therapy sessions within each therapy group. This is in contrast
to the analytically much simpler scenario of ``closed''
enrollment groups, in which the same set of clients is expected
to attend the same set of sessions with no change in
membership, such that the correlation in the outcomes for
clients in the same closed group could be modeled using a
hierarchical model with random terms for each closed therapy
group.

\subsection{Previous approaches to modeling data from rolling
therapy groups}

Despite the ubiquity of groups with rolling admissions in
practice
 [\citet{kadd2001}; \citet{kad2004}; \citet{mon2002}; \citeauthor{Rohs2001} (\citeyear{Rohs2001,Roh004});
 \citet{davi2005}; \citet{gran2005}; \citet{csat2005}],
very little attention has been devoted to developing
appropriate statistical methods for such data. In a review
article, the possibility was discussed of using standard
hierarchical models that used \textit{group} as the clustering
variable (e.g., Figure~\ref{fig1a}), but was dismissed because it ignores the complex
attendance pattern \textit{within} the group [\citet{morgfals2006}].
Another possibility discussed therein was to break up
the group into subgroups in some fashion and use a multiple
membership model, but the independence assumption typically
invoked for such models was criticized. Perhaps even more
frequently, the approach taken in therapy group studies---with
rolling or closed groups---is to ignore the group structure
altogether [\citet{leekthom005}; \citet{robe2005}; \citet{bauesterhall008}].

To date, the only analytic method specifically developed and
applied to data collected from clients as they attend rolling
group sessions did not explicitly model the correlation among
outcomes from clients within a group [\citet{morgals2007}].
Rather, the authors of that study fit models assuming no
correlation but adjusted the standard errors to account for
possible correlation among clients who attended the same {\it
group} (not \textit{session}) using a robust standard error
(sandwich) estimator [\citet{LianZegelong1986}]. The authors
also addressed the potential problem of nonignorable missing
data [e.g., Little (\citeyear{Littmode1995})] from clients who
chose not to attend all sessions by using a pattern mixture
model, where patterns depended on latent classes related to
number of sessions attended and client characteristics. \citet{morgfals2008b} found that modeling
different missing data (attendance) patterns helped with variance estimation, in that nominal Type 1 error rates were preserved.

However, there are some important limitations to this approach.
First, even when nonignorably missing data are a major
problem, modeling missing data patterns does not account for
the correlation structure depicted in
Figures~\ref{fig1a} and \ref{fig1b}, leading to concerns that even
pattern-mixture modeling without accounting for the intra- or
inter-session correlation of client outcomes could lead to
underestimated variances. Second, using the sandwich estimator
is only appropriate in studies with large numbers of rolling
groups. The sandwich estimator will only provide a consistent
estimate of standard errors of parameter estimates, as the
number of therapy groups grows arbitrarily large, and it can be
biased when the number of groups is small
[\citet{LianZegelong1986}; \citet{Bell2002}]. The number of
rolling groups is small (as few as 1--4) in many rolling group
studies
[\citet{kadd2001}; \citet{kad2004}; \citeauthor{Rohs2001} (\citeyear{Rohs2001,Roh004})];
in the motivating BRIGHT study, there were just four rolling
groups. Third, the sandwich estimator only adjusts the standard
errors; it does not model the clustering of client outcomes
both within and across \textit{sessions} and does not provide
information about the sources of variance. Finally, even in
studies with large numbers of rolling groups, relying on the
sandwich estimator to adjust for correlation at the \textit{group}
level would be inefficient relative to model-based approaches
to directly model the correlation structure attributable to
\textit{session} attendance [\citet{firt1992}].

\subsection{Conditionally autoregressive priors}

Given these limitations, we pre\-sent a novel application of
conditionally autoregressive (CAR) priors
[\citet{Besa1991}] for modeling session-level
random effects to capture the interrelatedness of client
outcomes. CAR priors offer promise for adeptly modeling
correlations at the session level that are induced by session
attendance patterns. Though a quick glance at
Figures~\ref{fig1a}--\ref{fig1b} might suggest that a
potentially simpler time series approach might be appropriate,
the CAR prior can accommodate alternative scenarios that
frequently occur in practice, such as sessions offered at
unequally spaced points in time, multiple independent sets of
therapy group sessions or ``islands''
[\citet{Hodg2003}], and flexible definitions of
closeness of sessions that cover a broad range of complex
scenarios, such as accounting for number of clients shared by
sessions or variable lengths of time between sessions.

In this paper we describe the motivating BRIGHT study
(Section~\ref{motivatingstudy}) and present a hierarchical
Bayesian CAR model for the BRIGHT data
(Section~\ref{statmodel}). In Section~\ref{comparison} we
examine the relative performance of the Bayesian CAR approach
versus competing approaches for accounting for the
interdependence of client outcomes. We compare these approaches
using BRIGHT data to examine whether depressive symptoms
decrease during the course of group therapy in
Section~\ref{analysis}. We conclude by discussing the
implications of our method for therapy group data analysis in
Section~\ref{discussion}.

\section{Motivating study: Building Recovery by Improving Goals, Habits and Thoughts
(BRIGHT)} \label{motivatingstudy}

The BRIGHT study addressed the question of whe\-ther group
cognitive behavioral therapy (CBT) improves depressive symptoms
when delivered by substance abuse treatment counselors. The
goal of the group CBT offered in BRIGHT was to help persons
with depressive symptoms manage their depression and feel
better. The study occurred at four treatment sites that are
part of the Behavioral Health System (BHS) of Los Angeles
County, California. BHS is a nonprofit treatment provider and
is among the largest publicly funded programs in Los Angeles
County.

BHS clients were screened for study eligibility using a
two-stage process. First, BHS staff screened clients using the
Patient Health Questionnaire [\mbox{PHQ-9}; \citet{kroespit2002}]
14 days after entering residential treatment. The PHQ-9 is a
nine-item self-report measure that assesses the nine depression
symptoms from the DSM-IV depression criteria. If clients
received a PHQ-9 score indicating mild-to-severe depressive
symptoms ({i.e.,} \mbox{PHQ-9}${}\ge 5$), they\vadjust{\goodbreak} were asked whether
they were interested in being contacted to learn more about the
study. A~second screening was then conducted to verify that
clients had persistent depressive symptoms and thus were
eligible for the study by examining whether a second depressive
symptom score, the Beck Depression Inventory-II
[BDI-II; \citet{becbrow1996}], indicated
mild-to-severe depressive symptoms (BDI-II${}> 17$). In
addition, clients with bipolar disorder, schizophrenia, or
cognitive impairment were ineligible for the study, but clients
taking psychotropic medications for other reasons were eligible
so long as they had depressive symptoms as indicated by the
two-stage screening procedure. Participants were enrolled in
the study 3--4 weeks after admission to residential AOD
treatment. Seventy-six percent of those entering residential
AOD treatment were screened by BHS for depression and 25\% of
those clients were eligible for the study based on the second
screening. For this study, we examined data collected from
clients assigned to the group CBT intervention ($n=140$) and
analyzed changes in depressive symptoms during the course of
group CBT for the $132$ clients who attended at least one group
CBT session.

The BRIGHT group therapy consisted of $16$ sessions of group
CBT offered over an eight-week period that were organized into
four modules of four sessions each. Each module focused on the
relationship between depression and a particular aspect of a
person's life: Thoughts, Activities, People, and Substance Use.
Research suggests that each module is independently efficacious
[\citet{zeis1979}] and does not depend on the material
presented in the previous module, thus making the rolling
admissions policy reasonable. The four modules over $16$
sessions were offered on a rotating basis, as shown for one
rolling group in Figure~\ref{fig1b}. The CBT group had a
rolling admissions policy, as client composition of the group
was allowed to change every two weeks. Specifically, clients
were able to initiate treatment at the first session of any
module, which is graphically depicted for one rolling CBT group
in BRIGHT in Figure~\ref{fig1b}. In session one of each module
the cognitive--behavioral treatment model was described and
information about depression and its symptoms were provided to
the clients.

In all, $245$ group CBT sessions were offered to clients from
all four treatment sites. This included $14$ offerings of the
$16$-session sequence ($224$ sessions), $20$~additional
sessions from the ``Thoughts'' and ``Activities'' modules to
increase exposure to group CBT for those who joined the rolling
group late for three particular $16$-week sequences, and one
additional session that followed a long holiday weekend to make
up for poor attendance at the regularly scheduled session.
These $245$~sessions were divided into four CBT therapy groups
having distinct clients; the number of sessions for each of
these four groups was 36, 40, 40, and 129.

\begin{figure*}

\includegraphics{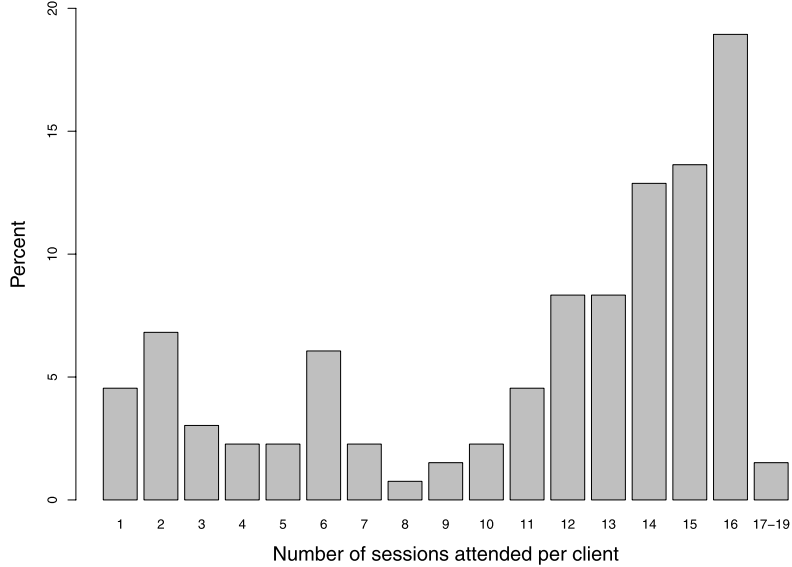}

\caption{Distribution of the number of sessions attended per client.}
\label{fig2b}
\end{figure*}

The number of sessions attended per client skewed toward more
versus fewer sessions (Figure~\ref{fig2b}). Seventy-three
percent of clients attended at least half of the sessions
(eight sessions), and $45\%$ attended $13$ or more sessions.
Sixty-three percent of clients attended at least one session
per module.

\section{Hierarchical Bayesian modeling of correlated therapy group session effects} \label{statmodel}

The goal of analyzing the client-level PHQ data collected from
clients during the course of group CBT is to estimate
client-level change over time. The core analytic model for this
is a growth curve model [or longitudinal growth model
(LGM) of Laird and Ware (\citeyear{Lair1982})], where there are
$i=1,\ldots,n$ clients in the study,  $s=1,\ldots,S$ therapy
group sessions offered over the course of the study, and
$t_{is}$ is the time that has passed since client~$i$ entered
the therapy group when attending session $s$. Note the
difference between $s$ and $t$: $s$ indexes all $S$ sessions
offered to all clients, while $t$ ranges from $0$ to the
maximum length of time between treatment entry and completion
at the client level; in BRIGHT, $t$ ranges from $0$ to $8$
weeks since entry into group. The repeatedly measured PHQ-9
score for client $i$ at session $s$ after attending the group
for time $t_{is}$ is denoted by $y_{is}$, and is modeled as
\begin{equation}\label{eq1}
y_{is} = \beta_0 + \beta_1 t_{is} + \sum_{k=1}^{K}\beta_{k+1}X_{ik} + b_{0i} + b_{1i}t_{is}+ \gamma_s + \varepsilon_{is},
\end{equation}
where baseline covariates, $X_1,\ldots, X_K$, have coefficients
$\beta_{2}$ through $\beta_{K+1}$; $\beta_0$ is the fixed
intercept term; $\beta_1$ the mean (fixed effect) rate of
change in $y$ for all clients; the random effects growth
parameters for client $i$ have distributions $b_{0i} \sim
N(0,\sigma^2_0)$ and $b_{1i} \sim N(0,\sigma^2_1)$;
$\varepsilon_{is}$ is the observation error with distribution
$N(0,\sigma^2_{\varepsilon})$; and the session-specific random
effects, $\gamma_s$, are included in the model to capture
session-specific variation in client outcomes. The
session-level random effects, $\gamma_s$, are modeled using a
convolution prior [\citet{Besa1991}]:
\begin{equation}\label{convolution-a}
\gamma_s = u_s + \nu_s,
\end{equation}
modeling the random effect of session $s$ as having both
structured and unstructured components to accommodate not only
structured, systematic correlation among sessions that is
expected given overlap client attendance patterns ($u_s$), but also
allows for independent, idiosyncratic session effects ($\nu_s$), allowing
for greater session-level variance than only a structured
component would offer. Thus, session $s$'s $(s=1,\ldots,S)$ effect is the sum of
 unstructured effects that are independently distributed:
 \begin{equation} \label{convolution-indep}
 \nu_s \sim N(0,\sigma^2_{\nu}),
\end{equation}
plus an effect that reflects
correlation structure among sessions, $u_s$'s, for which a CAR
prior is assumed:
\begin{equation}\label{convolution-b}
\mathbf{u} \propto \delta^{-(S-G)/2}\operatorname{exp}\biggl\{\frac{1}{2\delta}\mathbf{u}'\mathbf{B}\mathbf{u}\biggr\},
\qquad \mathbf{u}=(u_1,\ldots,u_S)',
\end{equation}
where $\mathbf{B}$ is a known $S$-by-$S$ matrix such that
$B^{-1}_{sj}=-w_{sj}$ and $B^{-1}_{ss} = \sum_j w_{sj}=w_{s+}$,
$w_{sj}$ is the $sj${th} element of a symmetric weight matrix
$\mathbf{W}$ that reflects the closeness between two
sessions and has diagonal elements $w_{ss}=0$ by definition,
and $G$ is the number of rolling therapy groups (or islands),
where therapy groups are composed of multiple, independent sets
of sessions such that clients attending sessions in one therapy
group do not attend sessions of another group.
Equation~(\ref{convolution-b}) implies a flat prior on each therapy
group's intercept, or fixed effect. To identify the therapy
group fixed effects and the session random effects, a
sum-to-zero constraint on each rolling therapy group's
$\mathbf{u}'s$ could be imposed. Since it is neither of
interest nor necessary to identify the therapy group fixed
effects for modeling the correlation of session effects, group
fixed effects are omitted from the model and are instead
implicitly included in $\mathbf{u}$ for model estimation, which
is similar to how \citet{Reic2006} treated
``island'' fixed effects in their application of CAR modeling to
disease mapping. These fixed effects could be obtained by
post-processing Markov Chain Monte Carlo (MCMC) output when
simulating conditional posterior distributions of model
parameters. One difference is that to model the intercept,
$\beta_0$, the $\mathbf{u}'s$ are centered about their
average across the $G$ therapy groups. Assuming covariates $X_k$
$(k=1,\ldots,K)$ are each centered about their means, this
parameterization implies that $\beta_0$ in this paper can be
interpreted the average outcome at~ti\-me~$0$ for clients
attending the average therapy group session, which avoids
estimating a trajectory (i.e., $\beta_0$ and $\beta_1$) that is
specific to a particular rolling group.

The mean of the distribution of the individual structured
session-level effect, $u_s$, is a weighted average of the
structured session-level effects of its neighbors:
\begin{eqnarray*} u_s | \mathbf{u}_{-s} \sim N
\biggl(\frac{\sum_{j\ne s} w_{sj}u_j}{w_{s+}},
\frac{\delta}{w_{s+}}\biggr)
\end{eqnarray*}
[\citet{Best2005}]. With this
in mind, we define two types of session closeness measures that
are of particular interest for rolling groups and guide the
weight matrix specification:
\begin{itemize}
\item   Closeness Type 1: sessions that are offered
    consecutively in time and in the same rolling group are
    neighbors: $w_{sj}=1$ if $|s-j|=1$ (where $j\ge 0$) and
    sessions $s$ and $j$ are in the same rolling group;
    $w_{sj}=0$ otherwise.
\item   Closeness Type 2: $w_{sj}$ reflects the proportion
    of clients in sessions $s$ and~$j$ that are common to
    both sessions.
\end{itemize}

Prior distributions are assumed for model parameters as
follows: $\beta_0 \propto 1$;  $\beta_1 \sim
N(0,\sigma^2_{\beta_1})$; $\bolds{\beta}_{2:k+1} \sim
N(0,\sigma^2_{\beta}\mathbf{I})$; $\sigma^{-2}_{\varepsilon}
\sim \operatorname{Gamma}(\psi_{y0}, \psi_{y1})$; $\sigma^{-2}_0
\sim\break
\operatorname{Gamma}(\psi_{00}, \psi_{01})$; $\sigma^{-2}_1 \sim
\operatorname{Gamma}(\psi_{10}, \psi_{11})$;
 $\sigma^{-2}_{\nu} \sim \operatorname{Gamma}(\psi_{\nu0}, \psi_{\nu1})$; and $\delta^{-1} \sim \operatorname{Gamma}(d_0, d_1)$,
  where $E(a)=a_1/a_2$ for a $\operatorname{Gamma}(a_1,a_2)$ specification.

\section{Comparison of alternative modeling approaches: Simulation study} \label{comparison}
$\!\!\!$We conducted a Monte Carlo simulation study to compare performance of alternative methods versus the hierarchical
Bayesian CAR approach for modeling rolling group data. We first outline our approach to generating the simulated data in
Section~\ref{41}, which is followed by a description in Section~\ref{42} of the analysis models fit to the simulated data
in order to conduct the model comparisons of interest, with results summarized in Section~\ref{43}.

\subsection{Simulated data generation} \label{41}
We used the BRIGHT study data to construct simulated data sets
with $245$ sessions attended by $132$ clients, each of whom
attended up to $16$ sessions, resulting in a total of $1473$
observations. All data were simulated to include session-level
random effects so that the performance of analytic models could
be compared under the presence of correlation among outcomes
among those attending common sessions in the rolling group.
Data were also simulated when assuming the absence versus
presence of missing data patterns to explore the performance of
competing models given the presence or absence of missing data
patterns.

The first set of data-generating models is given by
\begin{equation} \label{lgmcar}
y_{is} = \beta_0 + \beta_1t_{is} + b_{0i} + b_{1i}t_{is} +\gamma_s +   \varepsilon_{is},
\end{equation}
with the $\gamma_s$'s simulated as multivariate normally
distributed with mean vector $\mathbf{0}$ and covariance
matrix $\mathbf{G}$, where the correlation structure of
sessions is specified to be autoregressive with correlation
$\rho$ between adjacent sessions and the diagonals of
$\mathbf{G}$ equal to 1. We vary the degree of
autocorrelation among session-specific random effects,
$\gamma_s$, to reflect session-level autocorrelations of $\rho=
0.00, 0.25,$ and $0.50$; these values were selected  so the
average correlation among clusters approximated the range of
intra-class correlations (ICCs) reported for outcomes among
clients who share sessions in closed admission group therapy
studies [e.g., \citet{robe2005}; \citet{anderees2007}], since
data on correlation among outcomes for rolling groups are
unavailable. The simulation model parameters for the fixed
effect intercept and slope terms, $\beta_0$ and $\beta_1$, were
set at $15$ and $-0.5,$ respectively. Other simulation
parameters were set so that $(b_{0i}, b_{1i}) \sim
N(\mathbf{0},0.25\mathbf{I})$; $\varepsilon_{is} \sim
N(0,1);$ and $\gamma_s \sim N(0,1).$

The second set of data-generating models is a pattern-mixture model
(PMM) with two known missing data patterns to reflect whether
each client received at least 8 sessions $(n=96; 73\%)$, which
was regarded by BRIGHT study team members to be the minimum
adequate number of completed sessions and, thus, the research team wanted
to explore this distinction. The data-generating
PMM also included correlated session random effects,~%
$\gamma_s$:
\begin{eqnarray} \label{lcpmmcar}
y_{is} &=& \beta_0^{\ast} + \beta_1^{\ast}t_{is} + I_{\{R_i=1\}}\{\Delta_0 + \Delta_1t_{is}\} + b_{01} + b_{1i}t_{is} + \gamma_s +
\varepsilon_{is},\nonumber
\\[-8pt]\\[-8pt]
\qquad\ f(R_i|\pi_i) &=& \pi_i^{R_i} (1-\pi_i)^{1-R_i}. \nonumber
\end{eqnarray}
As before, $(b_{0i}, b_{1i})$ are the random growth parameters
for client $i$, $t_{is}$ is the time of observation of $y_{is},$ and
$\varepsilon_{is}$ is the observation-level error term. $R_i$ is
an indicator variable of the missing data pattern for client $i$;
$\beta_0^{\ast}$ and~$\beta_1^{\ast}$ represent the fixed
intercept and slope terms common for all clients, and
$\Delta_0$ and $\Delta_1$ the deviation from them,
respectively, for clients in missing data pattern $R=1$, so
that the marginal fixed intercept and slope equal $\beta_0 =
\beta_0^{\ast} + \pi\Delta_0$ and $\beta_1 = \beta_1^{\ast} +
\pi\Delta_1$. For the simulation study, $(\Delta_0, \Delta_1, \pi)$ are set to equal $(41.67,-1.83,0.273)$.

Prior distributions for most model parameters follow that
described in Section~\ref{statmodel}. Other priors are
$\Delta_0 \sim N(0,10)$, $\Delta_1 \sim N(0, 10),$
and $\psi_{y0}=\psi_{y1} =
\psi_{00}=\psi_{11}=\psi_{10}=\psi_{11}=1$, and we modeled the
prior probability of the missing data pattern, $\pi$, using a
$\operatorname{Beta}(1,1)$ prior. Since CAR model results can be sensitive to
hyperparameter choices for $\delta,$ we ran the simulation
study under two different hyperparameter choices for both the
prior on $\delta$ and for the precision of the unstructured
session effects, $\nu_s$:
\begin{eqnarray}\label{hyperprior1}
\psi_{\nu0}&=& \psi_{\nu1}=0.10, \qquad d_0=0.10,\qquad d_1=0.2,
 \\\label{hyperprior2}
\psi_{\nu0}&=&\psi_{\nu1}=1, \qquad d_0=0.5,\qquad d_1=0.0005.
    \end{eqnarray}
Under choice (\ref{hyperprior1}), the unstructured and
    structured variances are expected to have equal prior weight, since $\delta/\sigma^2_{\nu}$
    should be close to the average value of the sum of the weights for each
    session, which is two for this simulation study
    [\citet{moll1996}]. Choice (\ref{hyperprior1}) reflects a more
    informative yet sufficiently vague prior that could
    lead to better mixing of the Markov Chain Monte Carlo (MCMC) sampler [e.g., \citet{Reic2006}].
Choice~(\ref{hyperprior2}) reflects independent
    specification of the priors on the unstructured and
    structured components of the  session-level effects; hyperparameter values
    for $\psi_{\nu0}$ and $\psi_{\nu1}$ were selected as
    they would have been for a standard HLM analysis and
    values for $d_0$ and $d_1$ were used by \citet{kels1999} to place more prior mass near
    zero than choice (\ref{hyperprior1}), which can be helpful in
    situations with very little structure.

\subsection{Analysis models fit to simulated data}\label{42}
The analysis models we examined for the simulation study are as follows:
\begin{enumerate}
\item LGM, which is equation~(\ref{lgmcar}) but with $\gamma_s$'s set equal to zero.
This model reflects a widespread practice where no attempt is made to
    account for correlation among clients attending common
    sessions
    [\citet{leekthom005}; \citet{robe2005}]. Client-level
    random effects, $b_{0i}$ and~$b_{1i}$, are assumed to
    be independent across clients.
\item HLM, which is depicted by equation~(\ref{lgmcar}) but with
    $\gamma_s$'s independent and identically distributed. This reflects an approach
    that could easily be implemented in most standard
    statistical software packages but has been criticized
    for its independence assumption [\citet{morgfals2006}].
\item CAR, which is depicted by equation~(\ref{lgmcar}) but with $\gamma_s$ in equations~(\ref{convolution-a})--(\ref{convolution-b}), with the weights
    reflecting Closeness Type 1. Specifically, this
    specifies an autocorrelation, $\rho,$ between session
    effects for those sessions that are adjacent in time
    and belong to the same group, with
$\rho$ captured in the matrix, $B,$ in
equation~(\ref{convolution-b}).
\item  PMM, which is equation~(\ref{lcpmmcar}) but with $\gamma_s$'s set equal to zero.
 This model speaks to previous literature emphasizing the importance of modeling missing
 data classes for analyses of client data collected during the course of group therapy under
 rolling admissions [\citet{morgals2007}; \citet{morgfals2008b}].
\item CAR$+$PMM, which is equation~(\ref{lcpmmcar}) combined with
    equations~(\ref{convolution-a})--(\ref{convolution-b}) to model
    $\gamma_s$, with the weights reflecting Closeness Type~1. This approach will allow us to directly compare the
    PMM to a model with missing data patterns but that also
    allows for correlated session-level random effects to
    understand the relative effect of the missing data
    patterns on estimation.
\end{enumerate}
  Each analysis model was fit to data generated under every scenario, regardless of whether or not the generation
   and analysis models were congruent, which allowed for studying the robustness of models to misspecification.
   Conditional posterior distributions were simulated using MCMC as implemented in WinBUGS 1.4.3 [\citet{Lunn2000}]. We compared analysis
models based on how well the true values of $\beta_0$ and
$\beta_1$ were recovered. The standard deviations of these
parameters under the analysis models were also compared
[$\mathit{SD}(\beta_0)$ and $\mathit{SD}(\beta_1)$], along with posterior mean
deviance $(\mathit{Dbar})$ statistics to reflect goodness of fit. Since
we are interested in how well $y$ can be modeled given the
missing data pattern (for PMM and CAR$+$PMM), $\mathit{Dbar}$ is presented
for the growth submodel of equation~(\ref{lcpmmcar}).
Though the  Deviance Information Criterion (DIC) would reflect both model fit and complexity, there
is ambiguity on how to interpret DIC in mixture models,
including the concern that DIC is sensitive to choice of model
parameterization [\citet{Spie2002}]; thus, we do not present it here. MCMC
convergence was monitored by examining Gelman--Rubin statistics
for two Markov chains having different starting values
[\citet{GelmRubinfe1992}].

\subsection{Simulation study results}\label{43}

Results are shown in Table~\ref{tab1} for hyperparameter choice
(\ref{hyperprior1}), with hyperparameter choice
(\ref{hyperprior2}) omitted because the comparative results
were identical. The analysis model that is expected to have the
best performance given its consistency with the data-generating
process is denoted by ``best model'' for each of the six
data-generating scenarios (a)--(f) presented in
Table~\ref{tab1}.

When CAR should be the best model (Table~\ref{tab1}a--c), the
posterior means of~$\beta_0$ and $\beta_1$ are very similar to
their true values. Posterior standard deviations (SDs) of
$\beta_0$ were smallest for LGM and PMM, with the true CAR
model and the CAR$+$PMM models having the largest standard
deviations for $\beta_0$, demonstrating the tendency of models that ignore the correlation among outcomes
due to session attendance to underestimate the true variability in the parameter estimates. The posterior SDs of $\beta_1$ were
essentially equal for LGM and HLM across all three scenarios,
with the SD of $\beta_1$ under CAR being only about 2--3\%
larger. However, the SDs of $\beta_1$ under PMM and CAR$+$PMM
were 16--18\% larger than that under the true CAR model,
indicating greater noise in the variance of the slope given the
erroneous pattern-mixture model assumptions.

The two analysis models that omitted session random effects,
LGM and PMM, had identical performance, being tied for having
the highest posterior mean deviance $(\mathit{Dbar})$ statistics for each
scenario, thus indicating relatively worse model fit than models with session random effects. The
CAR$+$PMM and CAR $\mathit{Dbar}$ values were identical. The differences among CAR, CAR$+$PMM, and HLM were within Monte Carlo error,
indicating relatively comparable performance. Despite the aforementioned
limitations of DIC for this model comparison given two of the
models use mixture modeling, we compared results based on using
$\mathit{Dbar}$ versus DIC to assess whether there was any change in
results when factoring in model complexity. $\mathit{Dbar}$ and standard
DIC values (not shown) almost always had consistent patterns
across models, so the results based on comparing $\mathit{Dbar}$ across
models hold when factoring in model complexity along with model
fit. The only exception was the comparison of HLM and CAR in
Table~\ref{tab1}c, for which DICs could be unambiguously compared, given neither model is a mixture model.
In this case, CAR's $\mathit{Dbar}$ value was $10$ units greater than that for HLM, yet DICs for both models were equal $(\mathit{DIC}=4533)$
given the lower effective number of parameters in
the CAR model, thus showing the CAR model's increased efficiency in terms of number of effective parameters when there are larger
correlations among session effects ($\rho=0.50)$. The key message of Table~\ref{tab1}a--c is that accounting for
correlation of clients at the \textit{session} level using random
effects modeling offered a clear improvement over ignoring the
session-level clustering for the LGM and PMM models, with the
differences most pronounced when comparing posterior SDs of
model parameters and model fit (i.e., better posterior mean deviance).

\begin{table}
\tabcolsep=-1pt
\caption{Posterior means and standard deviations (SDs) of $\beta_0$ (true value${} = 15$) and
$\beta_1$ (true value${} = -0.5$), along with posterior mean deviance $(\mathit{Dbar})$ for five analysis models and six
data-generating scenarios. Simulation study results averaged over 100 simulated data sets. Monte Carlo
 standard error (mcse) reported for each estimate in parentheses}
\label{tab1}
\begin{tabular*}{\tablewidth}{@{\extracolsep{4in minus 4in}}lccccc@{}}
 \hline
   \textbf{Analysis}   & {$\bolds{\mathit{mean}(\beta_0)}$} &{$\bolds{\mathit{SD}(\beta_0)}$} & {$\bolds{\mathit{mean}(\beta_1)}$} &
   {$\bolds{\mathit{SD}(\beta_1)}$}&  {$\bolds{\mathit{Dbar}}$} \\
\textbf{model}&\textbf{(mcse)}&\textbf{(mcse)}&\textbf{(mcse)}&\textbf{(mcse)}&\textbf{(mcse)}\\
 \hline
            \multicolumn{6}{@{}l}{(a) Best model: CAR ($\rho=0.00$)} \\
 LGM & 15.00 (0.01) & 0.0828 (0.0003) & $-$0.498 (0.005) & 0.0518 (0.0003) & 5205 (9)\phantom{0} \\
 HLM & 14.99 (0.01) & 0.1026 (0.0003) & $-$0.495 (0.005) & 0.0519 (0.0003) & 4169 (7)\phantom{0} \\
 CAR & 14.98 (0.01) & 0.1131 (0.0003) & $-$0.492 (0.005) & 0.0531 (0.0003) & 4177 (7)\phantom{0} \\
 PMM & 15.00 (0.01) & 0.0910 (0.0003) & $-$0.496 (0.007) & 0.0627 (0.0003) & 5205 (9)\phantom{0} \\
 CAR$+$PMM & 14.98 (0.01) & 0.1125 (0.0003) & $-$0.495 (0.006) & 0.0614 (0.0003) & 4177 (7)\phantom{0}
 \\[3pt]
            \multicolumn{6}{@{}l}{(b) Best model: CAR ($\rho=0.25$)} \\
 LGM & 15.01 (0.01) & 0.0863 (0.0005) & $-$0.508 (0.006) & 0.0522 (0.0003) & 5126 (9)\phantom{0} \\
 HLM & 15.01 (0.01) & 0.1017 (0.0003) & $-$0.509 (0.005) & 0.0523 (0.0003) & 4162 (7)\phantom{0} \\
 CAR & 15.00 (0.01) & 0.1125 (0.0004) & $-$0.505 (0.005) & 0.0537 (0.0003) & 4171 (7)\phantom{0} \\
 PMM & 15.01 (0.01) & 0.0940 (0.0005) & $-$0.501 (0.007) & 0.0627 (0.0003) & 5126 (9)\phantom{0} \\
  CAR$+$PMM & 15.00 (0.01) & 0.1118 (0.0003) & $-$0.508 (0.006) & 0.0620 (0.0003) & 4170 (7)\phantom{0} \\[3pt]
                 \multicolumn{6}{@{}l}{(c) Best model: CAR ($\rho=0.50$)} \\
  LGM & 15.00 (0.02) & 0.0928 (0.0007) & $-$0.494 (0.005) & 0.0516 (0.0003) & 5021 (9)\phantom{0} \\
  HLM & 15.01 (0.01) & 0.1011 (0.0004) & $-$0.495 (0.005) & 0.0515 (0.0003) & 4169 (6)\phantom{0} \\
  CAR & 15.01 (0.01) & 0.1132 (0.0004) & $-$0.494 (0.005) & 0.0534 (0.0003) & 4179 (6)\phantom{0} \\
  PMM & 15.01 (0.02) & 0.0999 (0.0006) & $-$0.489 (0.006) & 0.0617 (0.0003) & 5020 (9)\phantom{0} \\
  CAR$+$PMM & 15.02 (0.01) & 0.1114 (0.0004) & $-$0.496 (0.005) & 0.0615 (0.0003) & 4178 (6)\phantom{0} \\[3pt]
                 \multicolumn{6}{@{}l}{(d) Best model: CAR$+$PMM ($\rho=0.00$)} \\
  LGM & 14.82 (0.01) & 1.378 (0.0009) & $-$0.208 (0.006) & 0.0687 (0.0005) & 5236 (10) \\
  HLM & 14.87 (0.01) & 1.382 (0.0007) & $-$0.244 (0.005) & 0.0730 (0.0005) & 4201 (7)\phantom{0} \\
  CAR & 14.59 (0.01) & 1.387 (0.0008) & $-$0.268 (0.006) & 0.0797 (0.0005) & 4211 (7)\phantom{0} \\
  PMM & 15.11 (0.01) & 1.412 (0.0009) & $-$0.488 (0.006) & 0.0931 (0.0006) & 5197 (10) \\
  CAR$+$PMM & 15.09 (0.01) & 1.405 (0.0007) & $-$0.494 (0.006) & 0.0917 (0.0005) & 4178 (7)\phantom{0} \\[3pt]
                 \multicolumn{6}{@{}l}{(e) Best model: CAR$+$PMM ($\rho=0.25$)} \\
  LGM & 14.78 (0.01) & 1.379 (0.0010) & $-$0.209 (0.006) & 0.0701 (0.0006) & 5146 (9)\phantom{0} \\
  HLM & 14.83 (0.01) & 1.381 (0.0009) & $-$0.240 (0.006) & 0.0734 (0.0005) & 4196 (8)\phantom{0} \\
  CAR & 14.55 (0.01) & 1.390 (0.0009) & $-$0.273 (0.007) & 0.0827 (0.0006) & 4205 (7)\phantom{0} \\
  PMM & 15.06 (0.01) & 1.411 (0.0010) & $-$0.484 (0.007) & 0.0933 (0.0006) & 5116 (9)\phantom{0} \\
  CAR$+$PMM & 15.05 (0.01) & 1.404 (0.0008) & $-$0.492 (0.007) & 0.0921 (0.0006) & 4173 (7)\phantom{0} \\[3pt]
                  \multicolumn{6}{@{}l}{(f) Best model: CAR$+$PMM ($\rho=0.50$)} \\
  LGM & 14.84 (0.02) & 1.383 (0.0011) & $-$0.226 (0.006) & 0.0715 (0.0005) & 5054 (9)\phantom{0} \\
  HLM & 14.87 (0.02) & 1.384 (0.0009) & $-$0.251 (0.005) & 0.0741 (0.0005) & 4204 (6)\phantom{0} \\
  CAR & 14.63 (0.02) & 1.397 (0.0010) & $-$0.307 (0.006) & 0.0876 (0.0005) & 4211 (6)\phantom{0} \\
  PMM & 15.11 (0.02) & 1.413 (0.0011) & $-$0.499 (0.007) & 0.0939 (0.0006) & 5031 (9)\phantom{0} \\
  CAR$+$PMM & 15.09 (0.02) & 1.406 (0.0007) & $-$0.507 (0.006) & 0.0932 (0.0005) & 4174 (6)\phantom{0} \\
   \hline
\end{tabular*}
\end{table}

Table~\ref{tab1}d--f shows the results when there are two
missing data patterns present in the simulated data, along with
session random effects. The model expected to perform best,
CAR$+$PMM, clearly did so in terms of yielding the correct
posterior mean values of $\beta_0$ and $\beta_1$ and having the
lowest $\mathit{Dbar}$. As expected, the non-PMM-based models (LGM, HLM, and CAR) yield biased coefficients, with the posterior mean of
$\beta_0$ underestimated by about $0.2$ units, while underestimation of
$\beta_1$ was about $50\%$ smaller in
magnitude than the true value of $-0.5$. PMM and CAR$+$PMM were
comparable in terms of estimating the posterior means and SDs of~$\beta_0$ and $\beta_1$. The SDs of~$\beta_0$
and $\beta_1$ under the non-PMM models were underestimated relative to the true CAR$+$PMM model. However, it is
important to note that the much smaller $\mathit{Dbar}$ value for CAR$+$PMM versus PMM indicates the CAR$+$PMM model
overall better captures aspects of the data that are not captured by $\beta_0$ and $\beta_1$, such as session-level
variation. Along these lines, there is a clear bias--variance trade-off for HLM and CAR versus PMM: HLM and CAR models
have better model fit according to $\mathit{Dbar}$ due to better capturing session-to-session variation, despite yielding biased
 posterior mean estimates of $\beta_0$ and~$\beta_1$.

To conclude, this simulation study demonstrates the importance of accounting for session-level variation in rolling groups to
improve model fit. These results show that the decisions of whether to model session effects or missing data patterns can be
made separately, and that these analytic strategies could both be crucial in rolling group therapy data analyses. We examine
the implications of these results for our motivating study in the next section.

\section{Analysis of depressive symptom scores from the BRIGHT study}\label{analysis}

\subsection{Data and methods}
Our analysis of the BRIGHT data focused on assessing whether
depressive symptoms decreased during the course of group
therapy. We focused on an analysis of change in PHQ-9
depressive symptom scores for the $132$ clients
assigned to the group CBT condition who attended at least one
session of group CBT from whom data on depressive symptoms were
repeatedly collected during their tenure in the CBT group.
PHQ-9 scores can be used to describe the patient's symptoms in
one of five interpretive categories: none (0--4), mild (5--9),
moderate (10--14), moderately severe (15--19), and severe
(20--27). In order to minimize reporting burden on the client, the PHQ-9 was completed at every other session
starting with the first session of each module as well as the
last scheduled session for each client. Data on demographic
characteristics, depressive symptoms, AOD use, and related
measures were collected from all study participants at a
baseline interview conducted within four weeks of treatment
entry.

We assessed whether depressive symptoms decreased over time by
examining the marginal posterior distribution of the time
(weeks) coefficient $\beta_1$. PHQ-9 depressive symptoms scores
were modeled while controlling for pre-treatment client
characteristics that we thought might vary among clients at
treatment entry and might be related to initial depressive
symptoms: demographic characteristics (age; gender; and
race/ethnicity categorized as White/Caucasian, African
American, Hispanic, and other), physical and mental health
status using the Short Form-12 version 2
[SF-12v2; \citet{warurn2002}], alcohol
 and drug severity using the
Addiction Severity Index evaluation indices
[\citet{alte2001}], and client's self-reported desire
to quit score and abstinence goal using the Thought about
Abstinence scale [\citet{hal1990}]. We considered
adding treatment site to the model but did not do so since
outcomes did not significantly vary by site.

The following models were fit to these data:
\begin{itemize}
\item CAR [equations~(\ref{eq1})--(\ref{convolution-b})].

\item HLM [equation~(\ref{eq1}) assuming independent and identically distributed {(i.i.d.)}
session random effects: $\gamma_s \sim
N(0,\sigma^2_{\gamma})$].

\item LGM [equation~(\ref{eq1}) but with $\gamma_s$'s fixed to equal
zero].

\item PMM [equation~(\ref{lcpmmcar})
but with $\gamma_s$'s fixed to equal zero], and with two missing data patterns defined by having attended
fewer than eight sessions (shorter-stay clients) versus at least eight sessions (longer-stay clients);
since the BRIGHT developers believed this was the minimum adequate number of sessions, we chose the patterns to explore this
distinction.

\item CAR$+$PMM [equation~(\ref{lcpmmcar}) with $\gamma_s$ modeled as
equations~(\ref{convolution-a})--(\ref{convolution-b})].

\item HLM$+$PMM is examined as well [equation~(\ref{lcpmmcar}) with {i.i.d.} session random effects assumed:
 $\gamma_s \sim N(0,\sigma^2_{\gamma})$], given HLM's equivalent performance to CAR in Table~\ref{tab1}a--c.
\end{itemize}

The following values were assumed for hyperparameters:
$\sigma^2_{\beta_1}\sim \operatorname{Gamma}(1,\break 1),$ $\psi_{y0}=\psi_{y1}=1$,
$\psi_{00} = \psi_{01}=1$, $\psi_{10}=\psi_{11}=1$, $\pi \sim
\operatorname{Beta}(1,1)$, $\Delta_0 \sim\break \operatorname{Gamma}(1, 0.1)$, and $\Delta_1 \sim
\operatorname{Gamma}(1, 0.5)$. We examined two sets of hyperparameter choices
for the structured and unstructured variances
[equations~(\ref{hyperprior1})--(\ref{hyperprior2})].

CAR models were fit using weights for Session Closeness Types 1 and 2, as defined in
Section~\ref{statmodel}. The adequacy of the
linear growth assumption was confirmed by verifying that adding polynomial
terms for time into the model did not improve model fit.

In addition to concerns about potential nonignorable nonresponse due to early client
departure from group CBT that are modeled with the PMM, there was a second type of missing
data due to the fact that PHQ-9 scores were recorded at every other session. These intermittently
 missing-by-design PHQ-9 scores were imputed as missing at random (MAR) under the model described
 above using data augmentation to simulate the posterior predictive distributions of the missing PHQ-9
 scores [\citet{Schaanal1995}]. The MAR assumption is reasonable for this type of missingness, since
 it is due to the study design decision to minimize the
burden of repeatedly measuring depressive symptoms on AOD
treatment clients as opposed to the missingness being driven by choices made by study
participants that would have resulted in missing data
[\citet{Grah1996}].

Given that clients were only able to enroll in the group at the
first session of each module, we checked the sensitivity of our
results to this by repeating the analyses just described but
instead placing a CAR prior on module-level random effects and
omitting session-level random effects.

\begin{figure*}

\includegraphics{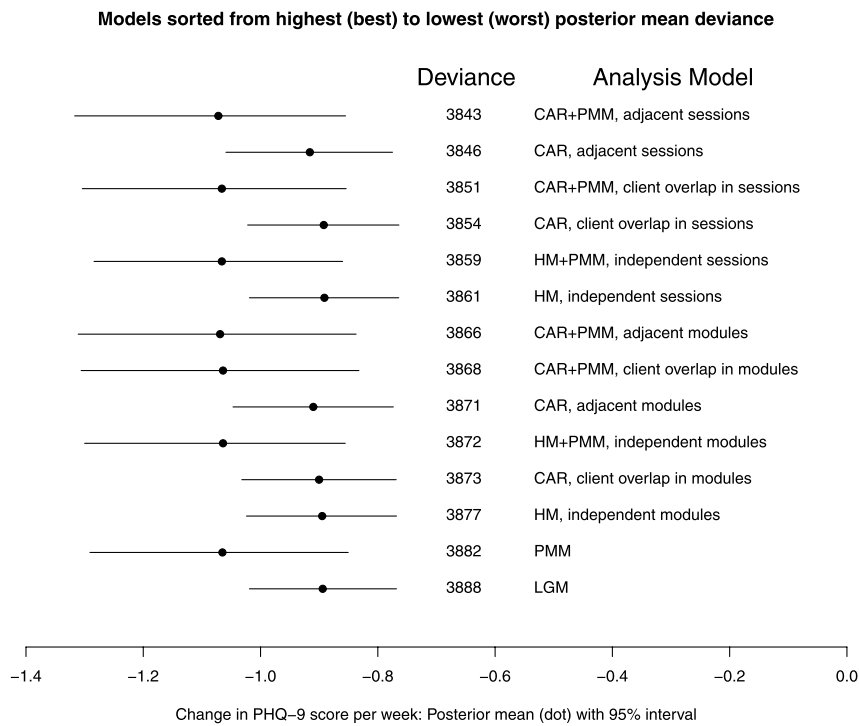}

\caption{95\% HPD interval for marginal slope parameter, along with posterior mean deviance and analysis model
description. Models sorted from lowest posterior mean deviance (indicating better model fit) to highest
(indicating worse fit). ``Closeness'' indicates closeness type for CAR priors and whether ``session'' or ``module'' indicates the clustering unit.}
\label{result}
\end{figure*}

\subsection{Results}
For each analysis model, Figure~\ref{result} displays its
$95\%$ highest posterior density (HPD) interval for the
marginal slope parameter, $\beta_1$, with results sorted by the posterior mean
deviance $(\mathit{Dbar})$, with the lowest values (indicating best model fit) at the top of the figure. For each analysis model that accounts for
the correlation of PHQ-9 scores among clients attending the
same therapy group, it is noted on Figure~\ref{result} whether
such correlation was modeled by assuming random effects for
either sessions or modules. For models employing CAR priors,
the closeness type used to define session (or module) distance
is also noted on Figure~\ref{result}, with ``adjacent'' referring to Closeness Type 1 and ``client overlap'' to Closeness Type 2.

Regardless of analysis model, depressive symptoms as measured
by \mbox{PHQ-9} decreased over time as clients participated in group
CBT, as indicated by the $95\%$ HPD intervals for the slope,
$\beta_1$, falling entirely below zero. The posterior mean
slopes displayed as dots in Figure~\ref{result} indicate the average
decrease in PHQ scores per week. For example, for the ``CAR,
Closeness~Ty\-pe~1'' model that includes random session effects,
the posterior mean decrease in PHQ scores was $-0.92$ points
per week; over the course of an expected eight-week
participation in group, PHQ scores would be expected to
decrease by $7.4$ points, which could decrease depressive
symptoms by at least one PHQ-9 depressive symptom severity
level category (see Section~\ref{motivatingstudy} for PHQ-9
category interpretation). These results were robust to the
choice of hyperparameter values examined, so results under just one
hyperparameter specification are reported.

Goodness of fit as measured by the posterior mean deviance
$(\mathit{Dbar})$ varied across the models (Figure~\ref{result}). The
LGM and PMM had relatively large posterior mean deviances,
indicating inferior performance to all other models considered
that explicitly modeled the correlation structure among
outcomes. The HLM models had worse fits than their respective
CAR models; for example, $\mathit{Dbar}$ for HLM with random session effects
was $3861$, versus $\mathit{Dbar}=3846$ and $3854$ for the two CAR
models with session effects. Models with session-level versus module-level random effects had better
goodness of fit, with the minimum difference in $\mathit{Dbar}$ between
any two otherwise similar models being $13$ points. For the CAR
models, Closeness Type 1 (``adjacent'') resulted in better model fit than
Closeness Type 2 (``client overlap''), with the difference more pronounced for
models with session random effects (change in $\mathit{Dbar}$ of 8
points) than module random effects (change in $\mathit{Dbar}$ of a
negligible $2$ points). Thus, for these data, the temporal order of
session better characterized
the interrelatedness of client outcomes than the measure of
client overlap in session attendance.

CAR$+$PMM and HLM$+$PMM resulted in models with posterior mean
deviances that were up to $5$ points lower than that of the corresponding CAR or
HLM models, respectively; however, this discrepancy shrinks to 2--3 points when focusing on the better-fitting (i.e.,
lower posterior mean deviance) models with session rather than module random effects. The contrast of these models to the relatively poor
goodness of fit of PMM only ($\mathit{Dbar}=3882$) emphasizes how critical it is to model
correlation in client outcomes at the session level regardless of whether one chooses to model missing data patterns. All PMM
models resulted in much wider HPD intervals than the analogous
non-PMM model; for example, the $95\%$ HPD interval of the
``CAR$+$PMM, adjacent sessions'' was $60\%$ wider than
that under ``CAR, adjacent sessions.'' These results demonstrate the dilemma of the PMM for this data set:
Allowing parameters to differ by missing data pattern can remove bias if the patterns differ, but it can add considerable
variance because of the small number of time points to estimate the slope for the short-stay group. In the BRIGHT example
the lack of improvement in fit suggests the non-PMM CAR model with more precise slope estimates is preferable for our main goal
of examining whether PHQ-9 scores decrease over time.

Despite our preference for the simpler CAR model for modeling the margi\-nal change in PHQ-9 over time,
examining PHQ-9 trajectories for shorter- versus longer-stay clients may be of interest for some research goals
[\citet{morgfals2008b}]. The posterior mean trajectory of \mbox{PHQ-9} scores for $27\%$ of clients who completed
 fewer than half of the sixteen planned sessions under the best-fitting CAR$+$PMM model had intercept of 17.75 and
 slope $-$1.62, while the trajectory for the longer-stay clients had intercept 14.74 and slope $-$0.86. The difference in the intercepts
 was 3.0 points [$95\%$ posterior probability interval: $(1.27, 4.77)$] and for the slope $-0.76$ [$95\%$ interval: $(-1.44, -0.06)$].
 Shorter-stay clients entered the group with more severe depressive symptoms on average but also made more rapid improvements during their
 stay. However, the projected PHQ-9 score after 4 weeks (i.e., after 8 consecutive sessions) for both shorter- and longer-stay clients was
 11.3 points, suggesting the more severely depressed shorter-stay clients caught up to the less severely depressed longer-stay clients after
 4 weeks. Greater rates of improvement among clients obtaining relatively fewer therapy sessions might suggest that clients stay in
 treatment until they reach a ``good enough'' level of improvement [\citet{Bald2006}], with the shorter-stay group achieving
 greater reductions in depressive symptoms but then leaving treatment early after achieving some symptom reduction.

The posterior summaries of the total session-level effects
(Figure~\ref{fig3}) highlight the distinct characterizations of
the interrelatedness of session random effects under CAR$+$PMM
with Closeness Type 1 [CAR$+$PMM(1); top row], CAR$+$PMM with
Closeness Type 2 [CAR$+$PMM(2); middle row], and HLM$+$PMM (bottom
row) for each of the four distinct rolling groups. The
CAR$+$PMM(1) model results, for which session closeness was
assumed to be of Type 1, show more variation in client outcomes
associated with the session-level random effects than when
client overlap itself defines the closeness of sessions
[CAR$+$PMM(2)]. Since client overlap largely occurred in
temporally adjacent sessions in this study, CAR$+$PMM(1) is a
more succinct way of summarizing client overlap than CAR$+$PMM(2)
in this particular analysis.

\section{Discussion}\label{discussion}
\begin{figure*}

\includegraphics{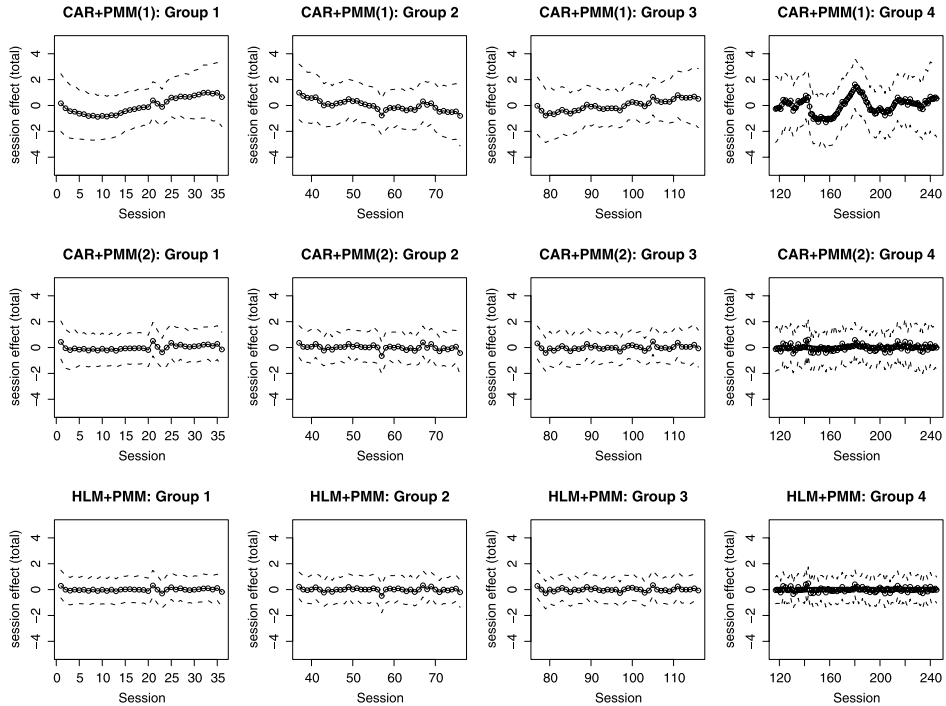}

\caption{Posterior means and $95\%$ HPD intervals for total session-level effect for the four distinct rolling therapy
groups under models CAR$+$PMM assuming Closeness Type 1 (\textup{top row}), CAR$+$PMM assuming Closeness Type 2
(\textup{middle row}), and HLM$+$PMM (\textup{bottom row}).}
\label{fig3}
\end{figure*}

We have presented a novel application of conditional
autoregressive modeling to address an important gap in the
literature on the analysis of data from rolling therapy groups.
Our modeling framework also more broadly addresses the
infrequent use of appropriate statistical methods for data from
therapy group studies, which even occurs for analytically more
straightforward closed admissions groups
[e.g., \citet{leekthom005}; \citet{robe2005}; \citet{bauesterhall008}]. Our approach avoids the limitations of
previously proposed strategies by directly modeling the
interrelatedness among client outcomes attributable to session
attendance rather than correcting for correlation at the therapy group
level. This makes our method applicable to the vast majority of
rolling group studies by not imposing an unrealistic
restriction that the number of rolling groups required for
analysis must be large. CAR modeling also provides a~better
understanding of the group dynamic and its effect on client
outcomes, as depicted by Figure~\ref{fig3} that shows the
pattern of session-level effects and their changes over time.
Our modeling approach is aligned with theoretical ideas about
how group therapy works by estimating a session-level
``invisible'' random component that can only be inferred from
client interactions [\citet{etti2002}].

The CAR model assumed for session-level correlations is
sufficiently flexible to cover a range of possibilities
encountered in practice, such as choosing appropriate session
closeness definitions that are most appropriate for a~given
therapy group. For example, the consecutive session closeness
measure (Closeness Type 1) would be appropriate for modeling
correlations between sessions when clients are expected to
attend consecutive sessions. However, the client overlap
(Closeness Type 2) measure might be more appealing when
expected or actual client attendance is irregular. For
situations in which clients attend multiple therapy groups that
focus on different issues---for example, one group for depression and
another for AOD use---the CAR-based framework could be readily
extended to model the closeness between sessions not just
within but also across therapy groups. Multivariate extensions
of the CAR prior could be applied to data on multivariate or
two-part outcomes, which are common in AOD treatment research
[\citet{Olse2001}; \citet{LiuMaJohnmult008}].

The CAR-based framework could be extended in other ways that
are relevant for the application. For example, we illustrated
how pattern-mixture modeling could be combined with our core
Bayesian hierarchical model using CAR priors to accommodate
concerns about nonignorably missing data due to early client
departure from treatment that are widespread in AOD treatment
studies such as BRIGHT. However, unlike \citet{morgfals2008b}, our simulation study
shows that it is not sufficient to only include missing data
patterns into longitudinal models of rolling
therapy group data: One must also directly model the correlation
structure of client outcomes induced by the rolling admissions
policies to fully capture the interrelatedness of client
participation in rolling therapy groups and make most efficient use of the data.
Finally, we are currently developing extensions of this methodology to
accommodate modeling of post-treatment outcomes in order to
fully understand the longer-term benefit of a rolling therapy
group versus an alternative treatment.

\section*{Acknowledgments}
We would like to thank Kimberly Hepner and Suzan\-ne Perry for their roles in data collection
and Annie Zhou for data processing.

\printaddresses

\end{document}